# Four methods for generation of turbulent phase screens: comparison

MIKHAIL CHARNOTSKII

*Erie, CO 805166, USA*
*\*Mikhail.Charnotskii@gmail.com*

**Abstract:** We introduce a new method for generation of the phase screen samples with arbitrary spatial spectrum: Sparse Spectrum with uniform wave vectors (SU). Similar to the known Sparse Spectrum (SS) technique, it uses trigonometric series with random discrete support on the wave vector plane, but, unlike the SS technique, the random wave vectors are uniformly distributed on the individual segments of the wave vector plane partition. We compare the accuracy and computational effectiveness of the SU technique with ubiquitous subharmonics complemented DFT method, SS method, and recently published, randomized DFT technique [J. Opt. Soc. Am. B **36**, 3249 (2019)]. SSU and SS algorithms generate unbiased samples, for screens with bounded phase variance, and show the superior computational effectiveness for one megapixel and larger screens.

## 1. Introduction

Computer-generated turbulent phase screens are widely used imaging, adaptive optics, beam propagation and optical communication studies. They are an essential part of the split-step method [1, 2] where multiple screens are used to represent the optical wave propagation through extended atmospheric turbulence. The accuracy and computational efficiency of the phase screen algorithms are crucial to the performance of wave propagation modeling. We refer to [3] for a comprehensive review of screen simulation techniques.

Currently, the most common algorithms for the turbulent phase screens generation use discrete Fourier transform (DFT) of the random spectral amplitudes with statistics matching, to some extent, the desired phase spectrum [4]. The DFT phase is complemented by subharmonic (SH) terms compensating for the lack of the low-frequency DFT components that are crucial for the wide inertial interval with characteristic power-law spectra, [5]. As a result, the number of the spectral components always exceed the number of the spatial points where phase is evaluated. The spectral support for the DFT-SH is discrete sample wise and in average, which leads to the biased phase samples.

One of the alternatives to the DFT-SH is the Sparse Spectrum (SS) technique [4, 6]. SS is based on the assumption that an individual phase sample contains only a limited number of spectral components, and both the amplitudes and wave vectors of these components are random. Number of the SS spectral components can be much smaller than the number of spatial points. SS has continuous, in average, support at the wave vector plane with ability to generate non-biased phase samples.

Recently, Paulson, Wu and Davis, (PWD) [7], proposed a new technique based on the randomized DFT-SH sampling. PWD combines the randomized sampling idea of the SS with computational efficiency of DFT, and, to some extent, can alleviate the low-frequency deficiency of DFT, even without invoking the SH corrections. PWD uses uniform wave vectors' distributions on rectangular segments at the wave vector plane and random complex amplitudes with compound distributions.

Here, we introduce a new, Sparse spectrum with Uniform distributions (SU) technique. SU is based on the observation that the PWD method allows a continuous coverage of the wave vector plane, but does not need to be confined to the rigid Cartesian partition. We combine the uniform wave vector distributions and compound amplitude distributions of spectral amplitudes with flexibility offered by the SS partitions to achieve a technique that uncouples the number of spectral components from the number of the spatial points, and is able to produce unbiased phase samples while using a modest number of spectral components.

The following Section reviews the DFT-SH, SS and PWD techniques, and provides numerical examples of their accuracy based on the Von Karman turbulence spectrum model. In Section 3 we introduce the SU technique and discuss its benefits and disadvantages. Section 4 discusses computational performance of all four techniques. Section 5 we discuss the accuracy criteria, convergence of the sample statistics and some issues related to the low-frequency divergence of model turbulence spectra.

## 2. Review of three screen simulation techniques

All phase simulation algorithms considered here use trigonometric series representation for the phase $\varphi(\mathbf{r})$

$$\psi(\mathbf{r}) = \sum_{n=1}^{N} a_n \exp(i\mathbf{k}_n \cdot \mathbf{r}), \quad \varphi_1(\mathbf{r}) = \mathrm{Re}[\psi(\mathbf{r})], \quad \varphi_2(\mathbf{r}) = \mathrm{Im}[\psi(\mathbf{r})]. \tag{1}$$

Here $\psi(\mathbf{r})$ is an auxiliary complex phase, $\mathbf{r} = (x, y)$ is the point where phase is evaluated, $a_n$ are random complex spectral amplitudes, $\mathbf{k}_n = (p_n, q_n)$ are the wave vectors of individual spectral components, and $N$ is the total number of spectral components used to represent the phase. Typically, a zero-mean circular statistics is used for the complex amplitudes $a_n$, when

$$\langle a_n \rangle = 0, \quad \langle a_n a_m \rangle = 0, \quad \langle a_n a_m^* \rangle = s_n \delta_{nm}. \tag{2}$$

In this case real and imaginary parts of the complex series in Eq. (1) produce two phase samples that are uncorrelated but have the same second-order statistics.

The difference between the four algorithms discussed here is in the choice of the number of components and statistics of the amplitudes and wave vectors. Here, as is common in the literature, with a possible exclusion of [6], we are concerned only by the second statistical moment of the phase. We choose it in the form of the structure function [8]

$$D(\mathbf{r}) = \langle [\varphi_1(\mathbf{R}+\mathbf{r}) - \varphi_1(\mathbf{R})]^2 \rangle = 2 \iint d^2\kappa \, \Phi(\boldsymbol{\kappa})[1 - \exp(i\boldsymbol{\kappa} \cdot \mathbf{r})], \tag{3}$$

where $\Phi(\boldsymbol{\kappa})$ is the spatial spectrum of phase. The goal of the phase simulation algorithms is to generate phase samples having sample-derived structure functions that are close to the target structure function given by Eq. (3).

We compare simulation techniques based on the popular Von Karman spectral model of turbulence. In application to the phase screens it implies that

$$\Phi(\boldsymbol{\kappa}) = \frac{C(\alpha) r_C^{-\alpha}}{\left(\kappa^2 + \kappa_0^2\right)^{1+\alpha/2}} \exp\left(-\frac{\kappa^2}{\kappa_m^2}\right), \quad C(\alpha) = \frac{\alpha 2^{\alpha-2} \Gamma(1+\alpha/2)}{\pi \Gamma(1-\alpha/2)}. \tag{4}$$

Here $\kappa_0 = 2\pi/L_0$, and $\kappa_m = 2\pi/l_0$, where $L_0$ and $l_0$ are the outer and inner scales of turbulence. Spectral normalization in Eq. (4) is explained by the fact that for zero inner scale, and infinite outer scale structure function is

$$D(\mathbf{r}) = \left(\frac{r}{r_C}\right)^{\alpha}, \tag{5}$$

and $r_C$ corresponds to the wave coherence radius for Kolmogorov turbulence. Parameter $\alpha$ typically is limited to $1 < \alpha < 2$. For numerical examples considered further we use $\alpha = 5/3$, $L_0 = 10m$, $l_0 = 1mm$, and $r_C = 1m$. Corresponding spectral parameters are $\kappa_0 \approx 0.63 rad/m$, and $\kappa_m \approx 6283 rad/m$. The linear size of the spatial grid is $L = 1m$.

It is important to distinguish the two types of differences between the desired, theoretical, structure function and one derived from a series of phase samples generated by a certain simulation technique. In case when there is a difference between the ensemble-averaged structure function and target structure function, the simulation algorithm is biased. Since any practical structure function estimate uses a finite number of samples, there is also a sampling error. In contrast to bias, sampling error can be reduced by increasing the number of samples.

*2.1 Discrete Fourier Transform complemented by subharmonics.*

### 2.1.1 Pure Discrete Fourier Transform method

For the case of the square spatial domain $-L/2 \leq x, y \leq L/2$, DFT-based method can be presented as

$$\psi_{DFT}(j\Delta x, l\Delta x) = \sum_{m,n=-N_{DFT}/2}^{N_{DFT}/2-1} a_{m,n} \exp\left(\frac{2i\pi}{N_{DFT}}(mj+nl)\right), \quad (6)$$

where $\Delta x = L/N_{DFT}$, $\Delta k = 2\pi/L$, and random complex amplitude have circular normal distribution with

$$\langle a_{m,n} \rangle = 0, \ \langle a_{m,n} a_{m',n'} \rangle = 0, \ \langle a_{m,n} a_{m',n'}^* \rangle = 2(\Delta k)^2 \Phi(m\Delta k, n\Delta k) \delta_{n,n'} \delta_{m,m'}. \quad (7)$$

Structure function of the DFT phase can be calculated from Eqs. (3, 6, 7) as

$$D_{DFT}(j\Delta x, l\Delta x) = 2(\Delta k)^2 \sum_{m,n=-N_{DFT}/2}^{N_{DFT}/2-1} \Phi(m\Delta k, n\Delta k)\left[1 - \exp\left(\frac{2i\pi}{N_{DFT}}(mj+nl)\right)\right]. \quad (8)$$

The main, and arguably only, advantage of the DFT method is that allows for the use of computationally efficient FFT algorithm. However, there are some serious drawbacks:
- Phase samples are periodic with period $N_{DFT}$, and practically useable spatial domain, at best, is limited to $-L/4 \leq x, y \leq L/4$.
- In the spatial domain phase is sampled only on a rectangular grid.
- Number of spectral components, $N_{DFT}^2$ is always equal to the number of points in the spatial domain, and can be very large.
- Phase spectrum is sampled only on a fixed rectangular grid. Consequentially, it is not possible, in general, to reproduce the required structure function, which, as clear from Eq. (3), requires continuous coverage of the wave vector plane.
- Cartesian wave vector grid inflicts some anisotropy on the phase samples, since frequency sampling rates are different along the coordinate axes and bisectors.
- Each term in Eq. (6) is responsible for contribution of the square patch

$$[\Delta k(m-1/2) \leq p \leq \Delta k(m+1/2) \cap \Delta k(n-1/2) \leq q \leq \Delta k(n+1/2)] \quad (9)$$

at the wave vector plane to the structure function, Eq. (3). However, in Eq. (6) this contribution is approximated by a simplest 2-D rectangle rule, causing more difference between $D(\mathbf{r})$ and $D_{DFT}(\mathbf{r})$.

- Specifically for the turbulent phase case, the (0, 0) term in the sum in Eq. (6) is set to zero, since it corresponds to the insignificant "piston" phase. This leaves the central square domain $|p|,|q| \leq \Delta k/2$ unaccounted for in the DFT structure function. Since low frequencies carry most of the energy in typical turbulent phase spectra this leads to substantial differences between $D(\mathbf{r})$ and $D_{DFT}(\mathbf{r})$.
- Arguably not so important for propagation simulation, the high-frequency content of DFT samples is limited to $k_{MAX}=\pi N_{DFT}/L$.

The last four items and Eq. (8) indicate that DFT is unable to generate the phase samples that have desired structure function. In other words, DFT phases are biased.

Dotted curves at the bottom of all three charts in Fig. 1 show structure functions of the DFT phase, Eq. (6) calculated based on 10,000 samples for $N_{DFT}=1024$, and $L=1\,m$. The lowest DFT frequency is $\Delta k \approx 6.3\,rad/m$, and the highest is $\Delta k\, N_{DFT}/2 \approx 3217\,rad/m$. Notably, the L- periodicity of the DFT phase samples results in $D_{DFT}(L)=0$, and DFT structure function is a very poor, biased, representation of the desired structure function. At least part of the reason for this discrepancy is the lack of low frequencies in the DFT model. Indeed, for this example the lowest DFT frequency is 10 times larger than $\kappa_0$.

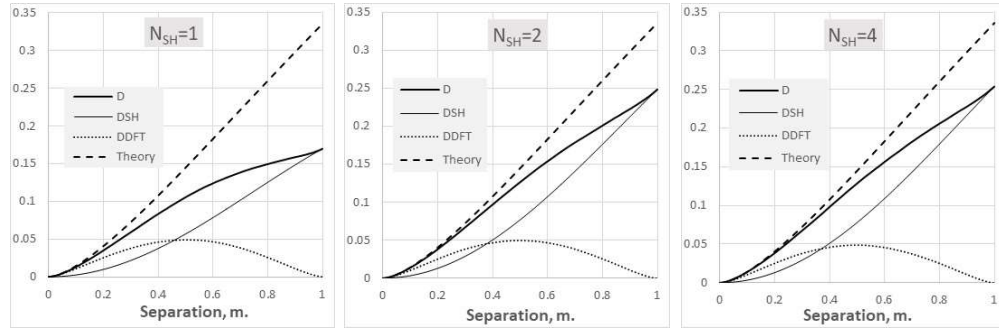

Fig. 1. Phase structure functions derived from DFT and DFT-SH phase samples. Dotted curve – $D_{DFT}(\mathbf{r})$, light solid curve - $D_{SH}(\mathbf{r})$, heavy solid curve – sum of the DFT and SH contributions. Dashed curve – target structure function, Eqs. (3, 4).

### 2.1.2 SH correction

Typically, deficiency of the DFT technique in application to the turbulent phase screens simulation is mitigated by adding to $\psi_{DFT}$ subharmonic spectral components $\psi_{SH}$ with support at the low frequency domain $|p|,|q| \leq \Delta k/2$. This can be done in many different ways, but most common SH correction [5] has the following form

$$\psi_{SH}(j\Delta x, l\Delta x) = \sum_{p=1}^{N_{SH}} \sum_{m,n=-1}^{1} a_{p,m,n} \exp\left(\frac{2i\pi}{3^p N_{DFT}}(mj+nl)\right)\left[1-\delta_{m,0}\delta_{n,0}\right], \quad (10)$$

where $N_{SH}$ is the order of SH correction, and SH random complex amplitudes $a_{p,m,n}$ are independent complex circular Gaussian variables with

$$\langle a_{p,m,n} \rangle = 0,\ \langle a_{p,m,n} a_{p,m',n'} \rangle = 0,\ \langle a_{p,m,n} a^*_{p',m',n'} \rangle = 2\left(\frac{\Delta k}{3^p}\right)^2 \Phi\left(\frac{m\Delta k}{3^p},\frac{n\Delta k}{3^p}\right)\delta_{p,p'}\delta_{n,n'}\delta_{m,m'}. \quad (11)$$

SH correction alleviates the low frequency deficiency of the DFT method, but still maintains some of the drawbacks of the pure DFT approach:

- Phase samples are still periodic with period $3^{N_{SH}} N_{DFT}$. At the first glance this allows for the unhindered use the full spatial domain $-L/2 \leq x, y \leq L/2$. However the high-frequency part of screens is still $N_{DFT}$ periodic. Certain wave propagation characteristics, e. g. scintillations, are mostly sensitive to the small-scale inhomogeneities. The effect of this periodicity on scintillation modeling is still to be investigated.
- In the spatial domain phase is still sampled only on a rectangular grid.
- Number of spectral components, $N_{DFT}^2 + 8N_{SH}$ is always larger than the number of points in the spatial domain, and can be very large for practical phase screens.
- Phase spectrum is still sampled only on a fixed discreet grid, and similar to the pure DFT case, it is not possible, in general, to reproduce the required structure function.
- Each term in combined Eqs. (6, 10) is responsible for contribution of a certain square patch at the wave vector plane to the structure function, and is still approximated by a 2-D rectangle rule, resulting in biased structure function
- The high-frequency content of phase samples is still limited to $k_{MAX} = \pi N_{DFT}/L$.
- FFT algorithm cannot be used for the SH component, Eq. (10). This can possibly offset the computational advantage of the FFT calculations.

Clearly, the DFT-SH technique should provide more accurate, but still biased phase samples. Fig. 1 shows SH components of structure function and final DFT-SH results for $N_{SH} = 1, 2, 4$. SH components make the greater contribution to the structure function than the DFT part, especially at larger separation, and improve the accuracy of simulated phase statistics. However, as evident from comparison of the . $N_{SH}=2$ and $N_{SH}=4$ cases there is no further improvement after $N_{SH}=2$. Indeed, for $N_{SH}=2$ the lowest wave number included in the simulation is $\Delta k/9 \approx 0.7 \, rad/m$, which is close to the $\kappa_0$, and there is not enough energy in the unsampled low-frequency domain to make a difference.

Remaining discrepancy between the desired and DFT-SH structure functions can be attributed to the insufficient high-frequency coverage, since the largest DFT wave number for $N_{DFT} = 1024$ is about $3217 \, rad/m$, while high-frequency cutoff is $\kappa_m \approx 6283 \, rad/m$. The left panel of Fig. 2 addresses this issue by showing the case of $N_{DFT} = 2048$, when highest DFT wave number is about $6434 \, rad/m$. There is no visible difference in the DFT components with the $N_{DFT} = 1024$ cases in Fig. 1, and as a result final DFT-SH structure functions remain same. This implies that the high-frequency part of phase is adequately represented by $N_{DFT} = 1024$. We conclude that the structure function difference should be attributed to the coarse sampling of the wave vector plane and simplified integration method mentioned earlier.

2.1.3 Frehlich correction

R. Frehlich made an attempt to improve the accuracy of the combined DFT-SS method [2] by modifying the variances of the SH spectral amplitudes. Instead of the rectangle rule, as in the last equation in Eq. (11) he used integrals over the spectral domains associated with spectral components. Namely [2] suggests that

$$\langle a_{p,m,n} a^*_{p',m',n'} \rangle = \int_{\frac{m\Delta k}{3^p}\left(m-\frac{1}{2}\right)}^{\frac{m\Delta k}{3^p}\left(m+\frac{1}{2}\right)} dp \int_{\frac{n\Delta k}{3^p}\left(m-\frac{1}{2}\right)}^{\frac{n\Delta k}{3^p}\left(m+\frac{1}{2}\right)} dq \, \Phi(p,q) \delta_{p,p'} \delta_{n,n'} \delta_{m,m'}. \qquad (12)$$

Frehlich correction requires additional calculations, of multiple 2-D integrals. However these calculations can be done outside the Monte-Carlo loop, and does not put a heavy computational burden on simulations. Central panel in Fig. 2 shows calculation results for $N_{DFT} = 1024$ and $N_{SH}=4$ with Frehlich correction. DFT component remains unchanged, as expected, but SH component increases noticeably, resulting in less than 10% overestimate of target structure function. Similar results were observed for $N_{SH}$= 2 and 8.

Clearly, Frehlich correction can be extended to the DFT variances in Eq. (7). Since number of DFT components significantly exceeds $N_{SH}$, computational burden can be substantial, but it is carried outside the Monte-Carlo loop. Right panel of Fig. 2 shows DFT-SH result with Frehlich correction applied both to DFT and SH parts. Noticeable increase of the DFT component is evident, and, combined with unaffected SH component, leads to significant overestimate of structure function. In our opinion, all this suggests, that Frehlich correction cannot be considered as a universal improvement for all spectral models.

Finally, we emphasize that all versions of the DFT-SH method provide biased estimates of the target structure function that, at best, can be asymptotically unbiased when both the number of spectral components and the size of the spatial domain are large.

## 2.2 Sparse Spectrum

Sparse spectrum (SS) technique for turbulence phase screen generation was proposed and analyzed in [3, 6]. The underlying physical concept is that each individual phase sample has random discrete support at the wave vector plane, and it is not necessary to use a large number of spectral components in order to accurately represent phase samples. Essential part of SS is that not only spectral amplitude, but also the wave vectors are random.

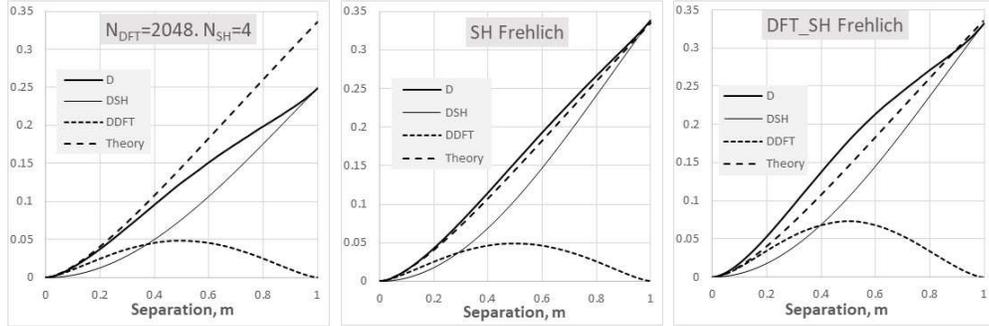

Fig. 2. Phase structure functions derived from DFT and DFT-SH phase samples. Dotted curve – $D_{DFT}(\mathbf{r})$, light solid curve - $D_{SH}(\mathbf{r})$, heavy solid curve – sum of the DFT and SH contributions. Dashed curve – theoretical structure function, Eqs. (3, 4).

Using the general form of Eqs. (1, 2) with $N = N_{SS}$, and assuming that probability distributions of wave vectors $\mathbf{k}_n$ are

$$P\{\mathbf{k}_n \in \mathbf{k} + d\mathbf{k}\} = p_n(\mathbf{k})d\mathbf{k}, \quad (13)$$

we calculate the SS phase structure function as

$$D_{SS}(\mathbf{r}) = \iint d^2k \sum_{n=1}^{N} s_n p_n(\mathbf{k})[1 - \cos(\mathbf{k} \cdot \mathbf{r})]. \quad (14)$$

Recalling Eq. (3), we conclude that the SS phase, as given by Eqs. (1, 2, 13) provides unbiased estimate of the desired structure function when

$$\sum_{n=1}^{N_{SS}} s_n p_n(\mathbf{k}) = 2\Phi(\mathbf{k}). \qquad (15)$$

SS phase has several advantages over the DFT-SH phase
- While individual samples of SS phase have discrete spectral support, ensemble averaged statistics has continuous spectral support, and spectrum of the SS phase is equal to the desired phase spectrum, $\Phi(\mathbf{k})$, [3]. In other words, SS phase samples are unbiased. This is not the case for any version of the DFT-SH phase.
- SS phase samples are not periodic, and it is possible to generate samples of any size, including long "phase ribbons" [3].
- Number of spectral components does not affect the accuracy of the structure function, as long as the full spectral support of $\Phi(\mathbf{k})$ is captured by joint support of $\{p_n(\mathbf{k})\}$. It is possible to account for a wide range of spatial frequencies with a moderate number of spectral components.
- There is no intrinsic limit on the high-frequency components. For simple, power-law spectral models it is possible to generate screens with zero inner scale [9].
- Location and number of spatial points is independent of the number of spectral components. This can be advantageous for propagating diverging or converging beams when screens rescaling is necessary, or for the ray tracing-based modeling of beam propagation, [10].

The only drawback of the SS technique is that it does not allow the use of the FFT algorithm, but requires direct series summation. However, as will be demonstrated further on, the moderate number of spectral component in the SS series makes it competitive to DFT in terms of computer time for several megapixel size phase screens.

For isotropic spectra $\Phi(k)$ it is natural to use polar coordinate form of $p_n(\mathbf{k})$ with uniform angular distribution of wave vectors directions

$$p_n(\mathbf{k}) d^2 k = p_n(k, \vartheta) k \, dk \, d\vartheta = P_n(k) dk \frac{d\vartheta}{2\pi}. \qquad (16)$$

We use a non-overlapping partition [6], of the desired range of wave numbers $[K_{MIN} < k < K_{MAX}]$, when

$$s_n = 2\pi \int_{K_{n-1}}^{K_n} k \, dk \, \Phi(k), \; P_n(k) = \begin{cases} k\Phi(k)\left[\int_{K_{n-1}}^{K_n} k \, dk \, \Phi(k)\right]^{-1}, & K_{n-1} \le k < K_n \\ 0, & k < K_{n-1} \cup k > K_n \end{cases}. \qquad (17)$$

Lognormal partition of wave numbers was identified as most efficient for turbulence phase screens modeling in [6], hence in Eq. (17) we choose

$$K_n = K_{MIN} \exp\left[\frac{n}{N_{SS}} \ln\left(\frac{K_{MAX}}{K_{MIN}}\right)\right]. \qquad (18)$$

Practical realization of random wave numbers with probability distributions $P_n(k)$ is relatively straightforward for pure power-law spectra, [6]. However for more complicated spectra, e. g. Von Karman spectrum, Eq. (4), it can become computationally expensive. We opted to use a simplified SS algorithm, which uses uniform distributions in each annulus $K_{n-1} \le k < K_n$. Namely we set

$$k_n = \sqrt{K_{n-1}^2 + \xi_n \left(K_n^2 - K_{n-1}^2\right)}, \qquad (19)$$

where $\xi_n$ are $U(0,1)$ i.i.d. random variables. This simplification introduces a small bias in the SS phase samples In case of the non-singular Von Karman spectrum, we chose $K_{MAX} = 2\kappa_m$ and $K_{MIN} = \kappa_0$, and complemented the partition, Eq. (18), by a zero term with $k_0$ uniformly distributed in the circle $k < K_{MIN}$. Numerical estimates showed that the final SS wave numbers range $0 < k < 2\kappa_m$ is sufficient to represent the support of the Von Karman spectrum. Namely sum of the complex amplitude variances $s_n$ was equal to the doubled phase variance for each $N_{SS}$ used.

Azimuthal angles are $N_{SS} + 1$ i.i.d. $U(0, 2\pi)$ random variables, and complex random spectral amplitudes are

$$a_n^{(SS)} = (\alpha_n + i\beta_n)\sqrt{s_n}, \qquad (20)$$

where $\alpha_n$ and $\beta_n$, are i.i.d. $N(0,1)$ random variables. In the case of Cartesian spatial grid when $\mathbf{r}_{j,l} = (x_j, y_l)$ it is convenient to calculate the $N_x \times N_y$ phase sample as a matrix product

$$\begin{aligned}\psi(x_j, y_l) = \mathbf{XAY}, \; A_{nm} = a_n \delta_{nm} \\ X_{jn} = \exp(ix_j k_n \cos\theta_n), \; Y_{ml} = \exp(ik_m \sin\theta_m y_l)\end{aligned}. \qquad (21)$$

Samples-derived SS structure functions are very close to the target structure function, and cannot be distinguished at the charts similar to Fig. 1 and Fig. 2. Figure 3 shows normalized RMS difference between the SS structure functions with different number of spectral components calculated from the different sample numbers of the SS phase screens and theoretical structure function, Eq. (3, 4).

$$\sigma_{SS} = \sqrt{\frac{1}{M}\sum_{i=1}^{M}\left(\frac{D_{SS}(r_i)}{D(r_i)} - 1\right)^2}, \qquad (22)$$

A total of $M = 100$ uniformly distributed separations $0 < r_i < 1m$ were used in Fig. 3. For $N_{SS} = 100, 200$ we observe biases of the SS screens that lead to at least 0.6% and 0.2% systematic errors. This bias is associated with the abovementioned replacement of the exact wave numbers distributions, Eq. (17) by uniform distributions, Eq. (19). However, for the larger number of SS components this bias, if even exists, is not substantial, not exceeding 0.1%. The residual error at the large number of samples probably should be attributed to accumulation of computer rounding errors. We conclude that 500 SS components is sufficient for the accurate SS screens simulations. It is worth noting that in case when the exact wave number PDF are used there should be no bias, and $\sigma_{SS}$ would be solely related to the finite number of samples.

*2.3 Paulson Wu and Davis technique*

Recently Paulson Wu and Davis [7] proposed a novel technique that uses randomized spectral sampling in the DFT method. Specifically, the complex phase is presented as

$$\psi_{PWD}(j\Delta x, l\Delta x) = \sum_{m,n=-N_{DFT}/2}^{N_{DFT}/2-1} a_{m,n}^{(PWD)} \exp\left(\frac{2i\pi}{N_{DFT}}\left((m+\xi)j + (n+\eta)l\right)\right) \qquad (23)$$

Here $\xi$ and $\eta$ are i.i.d. $U(-0.5, 0.5)$ random variables and complex spectral amplitudes are

$$a_{m,n}^{(PWD)} = (\alpha_{m,n} + i\beta_{m,n})\Delta k \sqrt{\Phi(\Delta k(m+\xi), \Delta k(n+\eta))}, \qquad (24)$$

where $\alpha_{m,n}$ and $\beta_{m,n}$, are i.i.d. $N(0,1)$ random variables. PWD spectral amplitudes have compound probability distribution. Conditional on $(\xi,\eta)$ distributions for $a_{m,n}^{(PWD)}$ are circular Gaussian, similar to DFT case, Eq. (7). However, the conditional second moment

$$\left\langle a_{m,n}^{(PWD)} a_{m',n'}^{(PWD)*} \right\rangle_{\alpha,\beta} = 2(\Delta k)^2 \Phi(\Delta k(m+\xi), \Delta k(n+\eta)) \qquad (25)$$

is random with probability distribution determined by the shape of the phase spectrum.

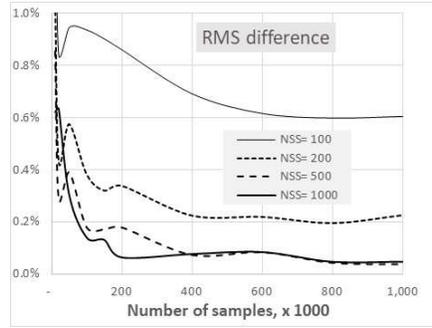

Fig. 3. Normalized RMS structure functions differences, Eq. (22) between the sample averaged SS structure functions and target structure function for different number of samples. Parameter is the number of the SS components $N_{SS}$.

It is crucial that the (0, 0) wave vector in the PWD series $(\Delta k\xi, \Delta k\eta)$ is uniformly distributed in the domain $|p|,|q| < \Delta k/2$ providing low frequencies input to PWD phase samples.

The benefits of the PWD technique are:
- With some modifications, [7], it allows the use of the FFT algorithm.
- Phase samples are not periodic.
- Low frequencies are represented in the phase samples.

However, certain drawbacks of the DFT method are carried over to the PWD
- In the spatial domain phase is still sampled only on a rectangular grid.
- Number of spectral components, $N_{PWD}^2$ is still large.
- The high-frequency content of phase samples is still limited to approximately $k_{MAX} = \pi N_{DFT}/L$.
- Generation of each random spectral coefficient requires calculation of spectral density at $N_{PWD}^2$ points inside the Monte-Carlo loop, Eq. (25). In the DFT case this is done outside the loop. For complicated spectral models its can add some computational time, offsetting the benefits of the FFT.

PWD found that a single spectral component in the low-frequency is not sufficiently accurate, and proposed "hybrid" technique by complementing the PWD phase, Eq. (23) by randomly sampled SH components

$$\psi_{SH}(j\Delta x, l\Delta x) = \sum_{p=1}^{N_{SH}} \sum_{m,n=-1}^{1} a_{p,m,n}^{(PWD)} \exp\left(\frac{2i\pi}{3^p N_{PWD}}\left((m+\xi_{p,m,n})j + (n+\eta_{p,m,n})l\right)\right)\left[1 - \delta_{m,0}\delta_{n,0}\right]$$
$$+ a_{N_{SH},0,0}^{(PWD)} \exp\left(\frac{2i\pi}{3^{N_{SH}} N_{PWD}}\left(\xi_{N_{SH},0,0}j + \eta_{N_{SH},0,0}l\right)\right).$$
(26)

Here $\xi_{p,m,n}$ and $\eta_{p,m,n}$ are i.i.d. $U(-0.5, 0.5)$ random variables and complex spectral amplitudes of sub-harmonics are

$$a_{m,n,p}^{(PWD)} = (\alpha_{m,n,p} + i\beta_{m,n,p})\frac{\Delta k}{3^p}\sqrt{\Phi\left(\frac{\Delta k}{3^p}(m+\xi_{m,n,p}), \frac{\Delta k}{3^p}(n+\eta_{m,n,p})\right)},$$ (27)

where $\alpha_{m,n,p}$ and $\beta_{m,n,p}$, are i.i.d. $N(0,1)$ random variables.

Hybrid algorithm populates low frequency spectral domain by $8N_{SH}$ points, that are independently and approximately log-uniform in wave number distributed and one more point that is uniformly distributed in the $\Delta k/3^{N_{SH}}$ by $\Delta k/3^{N_{SH}}$ square centered at zero frequency. This arrangement is very similar to the SS sampling with major differences being the square partitions of the wave vector plane and the compound distribution of spectral amplitudes.

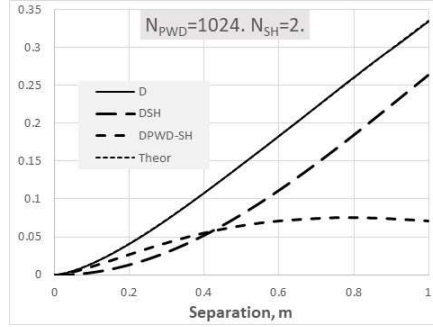

Fig. 4. PWD-SH sample-averaged structure function – heavy solid curve, and its components – dashed curves compared to theoretical expectation – dotted curve.

PWD failed to recognize [7] that, as is demonstrated in Appendix, their method is able to provide unbiased phase samples. Indeed, as long as the whole spectral support is captured by the pure PWD or PWD-SH frequency ranges, sample-based structure functions are very close to the target structure functions. Fig. 4 shows an example of the structure function calculated from 25,000 hybrid, PWD-SH samples, Eq. (23) and Eq. (26) as well as structure functions calculated separately from the PWD and SH components, and target structure function, Eq. (3, 4). Second-order PWD-SH algorithm accounts for the (0, 3217 rad/m) wave numbers range, and $\kappa_m = 6283$ rad/m for this example. Yet, uncaptured high-frequency range accounts for only a small fraction of spectral power, and target and sample-averaged structure functions are impossible to tell apart on the chart.

Figure 5 shows relative difference between the 25,000 samples-based PWD-SH structure functions with different orders of SH corrections and target structure functions. We observe no significant advantage of the SH samples. Similar conclusion can be drawn from the data presented in the Table 1 of [7] for $L_0 \leq 100\,m$. Parallel to the observations of [7], we note larger discrepancies at small separations that were attributed to the lack of high frequencies. Indeed, the highest frequencies in the PWD samples are approximately 3217 rad/m, while

$\kappa_m = 6283$ rad/m for all cases shown in Fig. 5. We examine the details of the PWD-SH accuracy issues in Discussion section.

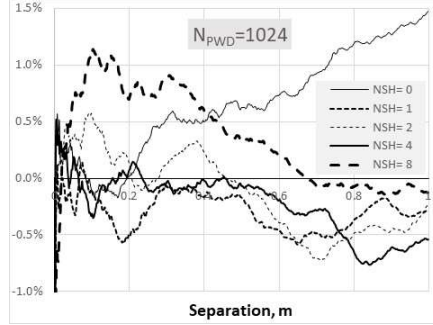

Fig. 5. Relative differences between the sample structure functions with different orders of SH components and the target structure function.

## 3. Sparse Uniform technique

In this section we introduce a new phase simulation method that uses Sparse Uniform (SU) random spectral sampling. SU combines some ideas of the SS and PWD techniques, and allows to generate unbiased phase samples while using a modest number of spectral components.

Calculation in Appendix, Eq. (A4) suggests that for any non-overlapping partition of the spectral support it is possibly to generate non-biased phase samples by using the uniformly distributed wave vectors in each subdomain and compound random amplitudes described by Eq. (A3). The partition, including the number of subdomains, is not connected with the number and location of the spatial points where phase is evaluated, and can have much less parts than the number of spatial points. Uniform distribution across each subdomain is the essential part of this technique, and we name it Sparse Uniform (SU) to emphasize these two features.

SU technique still uses the same trigonometric series form of Eq. (1). Similar to the SS technique, for isotropic spectra we use partition of the wave vectors plane in the concentric rings $K_{n-1} < k < K_n$, and polar coordinate representation of wave vectors $\mathbf{k}_n = k_n(\cos\theta_n, \sin\theta_n)$. After that, the SU phase is presented as

$$\psi_{SU}(x,y) = \sum_{n=1}^{N_{SU}} a_n^{(SU)} \exp(ik_n(x\cos\theta_n + y\sin\theta_n)). \quad (28)$$

Wave numbers $k_n$ are independent random variables, given by Eq. (19), and polar angles $\theta_n$ are i.i.d. $U(-\pi,\pi)$ random variables. Following the general case of Eq. (A3), complex amplitudes are calculated as

$$a_n^{(SU)} = (\alpha_n + i\beta_n)\sqrt{\pi(K_n^2 - K_{n-1}^2)\Phi(k_n)}, \quad (29)$$

where $\alpha_n$ and $\beta_n$, are i.i.d. $N(0,1)$ random variables.

Based on the analysis of [6], we use the same log-uniform partition of the $K_{MIN} < k < K_{MAX}$ annulus in $N_{SU}$ rings, Eq. (18), complemented by a circle $k < K_{MIN}$ as was done for the SS case. SU technique is unbiased and produces sample-averaged structure functions that are visually undistinguishable from the target when sufficient number of samples is used. Fig. 6 shows dependence of the RMS difference, Eq. (22) between the target and sample-averaged structure functions, as functions of the number of the SU samples used to estimate $D_{SU}$.

For all $N_{SU}$ values the $\sigma_{SU}$ decreases until about 200,000 – 400,000 samples, and remains stable after that at less than 0.1% level. We recognize the initial decline as the expected $\sqrt{N}$ sample error behavior, and attribute the small residual variance to the computational errors of numerical integration and summation of the hundreds of thousands of samples. Overall, we believe that Fig. 6 confirms the non-biased property of phase samples generated by our SU algorithm.

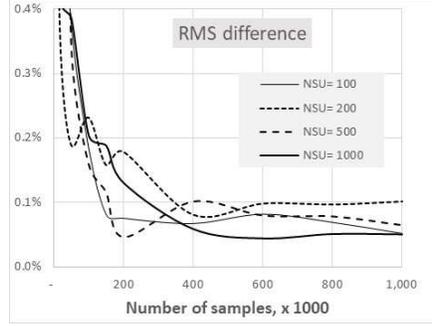

Fig. 6. RMS differences between the sample structure functions with different number of spectral components and target structure function.

## 4. Computational performance

The developments of the previous two sections indicate that three simulation techniques SS, PWH-SH and SU are all deliver unbiased phase samples, and the accuracy of the sample-averaged structure functions is limited mostly by a finite number of samples and computer accuracy. The only remaining issue for applications is the computational performance of these simulation techniques. We first compare performance of different techniques for the Von Karman spectral model, Eq. (4), and then discuss the effect of the spectral models on performance.

We estimated computational time per one phase screen by generating $N_S$ samples of square $M \times M$ complex phase screens by Matlab code and measuring the total computation time $T$ with `tic` and `toc` Matlab functions placed immediately before the start and after the end for the $N_S$-long Monte-Carlo loop. All preliminary calculations, such as the complex amplitude variances for the DFT and SS techniques were done outside the loop. Typical number of real phase samples $4000 \leq 2N_S \leq 20000$, and time per sample was calculated as $T/2N_S$. All calculations were performed on the same Windows 10 PC with I7-4770 CPU running at 3.4 GHz, and 12 GB RAM. We emphasize it is not the absolute time per sample in seconds, but the relative performance of different algorithms is crucial here.

Figure 7 shows computation time per screen for the SS and SU algorithm for different screen sizes, $M$, and different number of spectral components, $N_{SS}$. In both cases the same log-uniform partitions were used. Phase samples on square grids were calculated using matrix multiplication as described by Eq. (21). While SS algorithm does not require additional $N_{SU}$ calls to the spectral density function inside the Monte-Carlo loop, Eq. (17) cf. Eq. (29), there are no substantial differences between SS and SU techniques due to the relatively small number of spectral components. Dependences of computation time on screen size can be well approximated by power law with exponent in the (1.3 – 1.7) range with steeper slopes in for smaller number of components.

For SH terms both in DFT-SH and PWD-SH algorithms we arranged components of the SH wave vectors in two $(8N_{SH}+1) \times 1$ files, and used matrix multiplication similar to Eq. (21) to calculate the SH phase corrections. Fig. 8 shows relative increase of time-per-screen as a function of the number of added SH spectral components $8N_{SH}+1$ based on 5000 samples for each configuration. We observe that the addition of the SH components increases computation time by less than 20%. This is in sharp contrast to the data presented in Fig. 14a of [7], where for comparable cases at least 600% increase was reported.

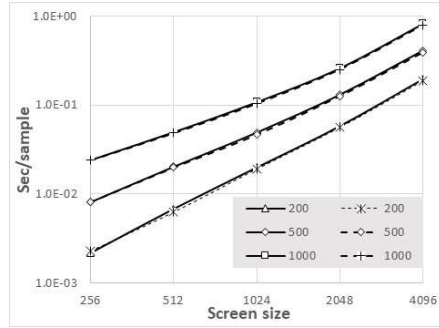

Fig. 7. Computation times per screen for the SS, solid curves, and SU, dashed curves, algorithms as functions of the screen size. Parameter is the number of SS components used.

Fig. 9 compares computation times for DFT-SH and PWD-SH algorithms. Following the earlier discussion, we observe that addition of the SH components has a minimal effect on the computation time. However the PWD computation time is approximately double the DFT time. This is caused by at least $M^2$ calls to the spectral density function inside the Monte-Carlo loop required by PWD, which are necessary for compound spectral amplitudes, Eqs. (24, 27). PWD [7] reported 25% - 30% increase of computation time, but it is not clear if the total or propagation simulation time or just screen generation time was used.

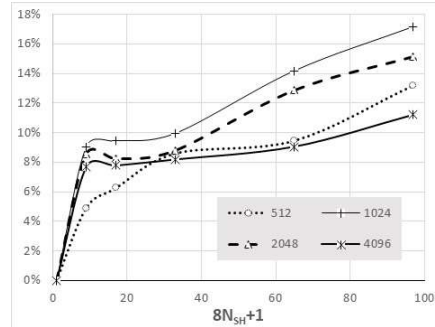

Fig. 8. Relative increase of computation times per screen for the PWD-SH phase screens with different number of the SH components as functions of the total number of spectral components added. Parameter is the screen size.

Fig. 10 compares performance of all four techniques with parameters chosen to provide a similar accuracy, as discussed in the previous sections. Namely, based on calculations presented in Fig. 3 and Fig. 6, we used 500 spectral components in the SS and SU cases, and fourth-order SH correction for PWD, based on results shown in Fig. 5. Both SS and SU algorithms have very similar performance, and they are faster than the PWD-SH algorithm for screens larger than 512x512. For the largest, 4096x4096 screens tested, SS and SU screens are more than three times faster. DFT-SH algorithm does not provide accuracy comparable to other techniques, Fig. 1, but is fastest for up to the 1024x1024 - size screens, due to the FFT computational advantage. However, for larger screens, restrained number of spectral components used in SS and SU overcomes their inability to use the FFT, resulting in about 50% disadvantage of DFT_SH in comparison to SS and SU.

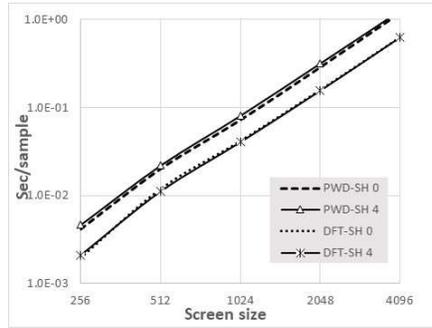

Fig. 9. Computation times per screen for the DFT-SH and PWD-SH algorithms as functions of
the screen size. Dashed curves – no subharmonics. Solid curves – fourth order subharmonics.

PWH-SH technique requires $M^2 + 8N_{SH}$ calls to the spectral density function from inside the Monte-Carlo loop. This can cause a noticeable computational burden for complicated spectral models. We illustrate this in Fig. 11, where ratios of computation times for three spectral models to computation times for the Von Karman mode are shown for the PWD algorithm with fourth-order SH correction. Three models are: band-limited 11/3 power law spectrum, [6], Andrews version of the Hill model, [11], and Nikishov & Nikishov sea water turbulence model [12]. Sea spectrum has the most complicated analytical form including multiple exponents and fractional power terms. This results in up to 70% increase in computation time relative to the simplest power-law spectrum. Calculations for the Von Karman model are predictably slower than for the power-law case, but the difference is insignificant.

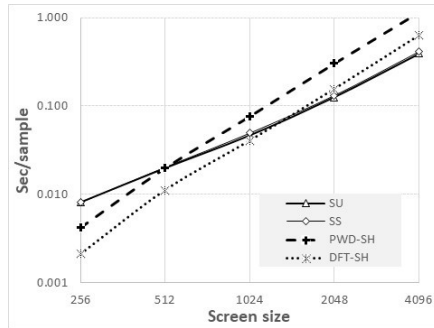

Fig. 10. Computation times per screen for the SS, PWD-SH, and SU algorithms with
comparable accuracy and DFT-SH as functions of the screen size. SS and SU algorithms used
500 spectral components, and PWD-SH used fourth-order SH.

## 5. Discussion

### 5.1 Non-Gaussian features

As was discussed in [6] and recently in [7], all three non-biased models produce non-Gaussian random fields. Indeed, unlike the classic DFT case, each term in the SS, SU or PWD series is not Gaussian due to the presence of a random wave vector. Certainly, it can be expected that a sum of many independent non-Gaussian terms would result in asymptotically Gaussian phases. However, this is true only when all terms in the series in Eq. (1) are comparable. As was noted in [7], even a simple change of outer scale affects probability distribution of phase.

There are at least two kinds of implications of the non-Gaussian phase. Turbulence wave propagation is based on the assumption of Gaussian refractive index fluctuations, and consequentially, the simple, geometrical optics phase used for the phase screens is Gaussian as well. Most of practically important parameters of propagating beam waves, mean irradiance, scintillation index, etc. are nonlinear functionals of the refractive index in the rigorous propagation theory and of the phase in the propagation modeling case. This means that phase screens having identical structure functions, but different probability distributions, as is the case for non-biased SS, SU and PWD-SH screens, will not provide identical statistics of propagating wave, and it is possible that none of these statistics will match theoretical results.

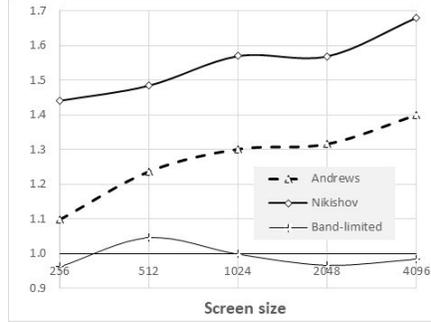

Fig. 11. Computation times per screen for PWD-SH algorithms as functions of the screen size for three model spectra.

Only wave statistics that are linear functionals of the turbulence spectrum are not sensitive to the phase probability, but only to the structure function. Unfortunately, [7] chose the perturbation theory result for angle of arrival variance as their accuracy test. Being an example of the linear dependence, this theory does not allow to examine the role of the non-Gaussianity of PWD-SH phase samples on the wave propagation simulations. In [3] the effect of the number of SS components on the Strehl number, and scintillation index in the image of the point source, both being nonlinear functionals of phase, was investigated and showed that $N_{SS} \geq 300$ is sufficient.

The second implication of the non-Gaussian distributions is related to the estimation of the accuracy of phase simulation techniques. Accepting that the normalized RMS phase difference, $\sigma_*$, as given by Eq. (22) for the SS case, is a suitable metric for all techniques, we note that it includes the $N_S$ sample-averaged structure function, and is, by itself, a random value. Strictly speaking, only in the case $N_S \to \infty$, or for ensemble-averaged case, it acquires statistically stable value. Unfortunately, for all three unbiased techniques this value is zero, and any non-zero estimates of $\sigma_*$ are just sample and/or computation noise. Fig. 3 gives examples of the RMS difference convergence for slightly biased case of simplified SS algorithm with small number of spectral components and Fig. 6 shows convergence to very small, computer accuracy-limited values for truly unbiased SU algorithm. We are not able to produce similar illustrations for the PWD-SH case, due to large computation time needed to produce at least 1 MP samples, but we can examine the normalized variance of the structure function estimate [6]

$$\sigma_D^2(\mathbf{r}) = \frac{\left\langle [\varphi(\mathbf{R}+\mathbf{r}) - \varphi(\mathbf{r})]^4 \right\rangle_{N_S}}{\left\langle [\varphi(\mathbf{R}+\mathbf{r}) - \varphi(\mathbf{r})]^2 \right\rangle_{N_S}^2} - 1. \qquad (30)$$

Here averaging is performed over a set of $N_S$ phase samples. This statistic provides the estimate of the rate of convergence of the sampled structure function to the theoretically expected one in terms of the relative error. At the same time it provides some insight in the probability distribution of phase. Namely, for Gaussian phase differences $\sigma_D^2(\mathbf{r}) = 2$. Figure 12 shows $\sigma_D^2(\mathbf{r})$ values for SS and SU samples based on 80,000 samples for each algorithm, at

100 uniformly distributed points. We used a log-uniform partition, Eq. (18), with a large $L_0 = 10^4 m$ outer scale in order to compare with the PWD case later. Smaller number of spectral components results in noticeable deviations from the Gaussian $\sigma_D^2(\mathbf{r}) = 2$, but for 500 components this fourth statistical moment matches the Gaussian case.

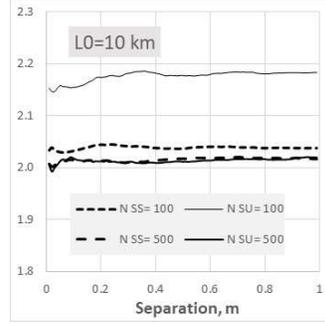

Fig. 12. Normalized variances of the structure function estimate for SS and SU techniques.

Authors of [7] came to conclusion that for larger outer scales it is necessary to add subharmonics in order "to achieve accurate RMS error statistics." We believe that this conclusion is based on the misinterpretation of their data in Table 1 of [7]. As we showed earlier, the PWD technique generates non-biased samples even without the SH corrections, as long as the high-frequency range is sufficiently captured. In other words, the RMS error tends to zero when an increasing number of samples is used for its estimation. However, for a finite number of samples, the sampling noise of the structure function estimates becomes critical, and the indeterminate dependence of RMS errors on $L_0$ and $N_{SH}$ in Table 1 of [7] is the evidence of this.

Figure 13 shows estimates of the $\sigma_D^2(\mathbf{r})$ for the full PWD-SH structure function and its DFT and SH components for $N_X = 1024$, $N_{SH} = 2$, and $L_0 = 10^3 m$. We observe that the variance of DFT component estimates is close to Gaussian for smaller separations, but increases twofold for the larger separations. This is due to the uniform partition of the wave vectors plane, where for higher wave numbers spectral density is almost constant over the $\Delta k \approx 6.3$ rad/m spans resulting in approximately Gaussian contribution to the phase differences at small separations. However at lower wave numbers spectral density changes substantially in each $\Delta k \times \Delta k$ subdomain, making phase differences at the large separations non-Gaussian. All 17 SH components are located close to the origin, but still at wave numbers much larger than $\kappa_0 \approx 0.0063$ rad/m, and have highly non-Gaussian distribution. Finally, total PWD-SH phase is non-Gaussian with $2 < \sigma_D^2(\mathbf{r}) < 9$ and growth with separation indicating that the small wave number components are responsible for the non-Gaussian behavior.

Increase of the order of the SH corrections populates the low wave numbers domain by multiple independent, but still non-Gaussian components. As a result, as shown in Fig. 14, $\sigma_D^2(\mathbf{r})$ becomes close to the Gaussian value of 2. Effect is similar for the larger outer scale values, but with higher SH orders needed to bring $\sigma_D^2(\mathbf{r})$ to the Gaussian value. We conclude that increase of the SH components does not affect the ensemble-averaged RMS accuracy, Eq. (30), but makes phase samples more similar to Gaussian, and accelerates convergence of samples-averaged structure function to the target. The former is beneficial for the propagation modeling, while the latter is irrelevant, in our opinion.

## 5.2 Outer scale issues

SU and PWD-SH can be directly applied to the power-law spectra with infinite outer scale. The wave vector uniformly distributed either in a small square or circle around origin will never hit singularity at zero. However, the large and erratic amplitude of a single randomly sampled low-frequency spectral component results in the high statistical variability of the structure function estimates based on a finite sample size. Fig. 15 shows 36 consecutive values of 1000 SU samples-based estimates of normalized RMS difference, $\sigma_{SU}$, Eq. (22). For a finite, 1000 m outer scale $\sigma_{SU}$ is subject to modest 50% variations, but for infinite outer scale we observe occasional 1000% spikes occurring when one or several lowest frequency components happen to be very close to the origin. While formally SU and PWD techniques are still unbiased for infinite outer scale, the intermittent variations of the RMS difference estimates makes it difficult to calculate a reliable estimate of the algorithms accuracy.

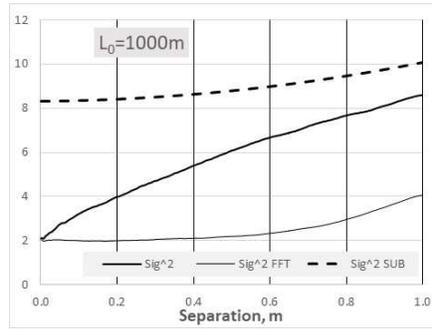

Fig. 13. Normalized variances of the structure function components for a second-order PWD-SH technique.

Phase screen with zero inner scale and power-law high-frequency behavior can be generated by SS technique [9], but require finite, albeit arbitrary large outer scale. In contrast, SU technique allows for zero inner scale, as described, but cannot handle the zero inner scale. It is straightforward to combine both techniques by introducing a separation wave number $K^*$ and using SU technique for $0 \leq k \leq K^*$ range and SS technique for $k \geq K^*$. This, hybrid technique is unbiased, but the accuracy estimation of sample-averaged structure functions can be problematic by the reasons discussed above.

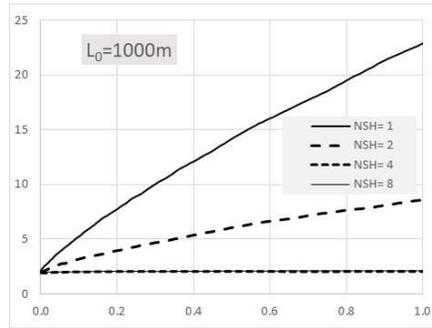

Fig. 14. Normalized variances of the structure function components different orders of SH corrections.

Accuracy criteria used in here and in [PWD] are based on the relative differences between the simulated and target structure functions. In the case when simulated phase is, at least approximately, Gaussian, structure function is the exhaustive statistics of the phase as a 2-D random field, and accurate replication of structure function warrants accurate statistics of any order. However, specifically in the optical phase case, relative error is not the best measure of the phase simulation quality. Phase has an intrinsic $[0,2\pi]$ range in optics. As a result, large, say 100%, phase errors relative to the 0.1 rad phase RMS difference at small separations are irrelevant in applications. However, small, say 1%, error relative to 300 rad RMS phase difference at large separations is not acceptable. This implies that accuracy of simulations models depends on the turbulence strength represented here by coherence radius and the typical spatial scales of propagating beam.

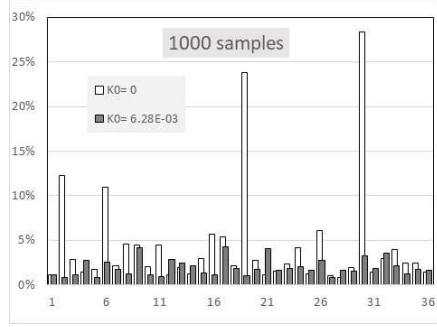

Fig. 15. Normalized RMS differences, $\sigma_{SU}$, Eq. (?), between 1000 SU samples estimates and target structure functions for infinite and 1000 m outer scales.

For SS and SU techniques we discussed only log-uniform non-overlapping partitions. However, SS and SU offer a great flexibility in the partitions choice. As one option, it is possible to use several independent spectral components in any segment of the wave vectors plane after adjusting the variances of spectral amplitudes, Eq. (17) and Eq. (29) accordingly. This can be used also in the SH correction terms of PWD-SH method in order to generate approximately Gaussian phases for low-order SH correction.

## 6. Summary

Here we discussed four computational technique for turbulent phase screens simulation and compared their accuracy and speed.

The classic, DFT method complemented by subharmonic components is most commonly used in wave propagation simulations, but as we showed, provides biased phase samples. It is only capable to generate phase samples restricted to a rectangular spatial grid, and uses a number of spectral component that exceeds the number of spatial points. As a result, and despite the advantages of the FFT algorithm, DFT-SH algorithm it is not the fastest for large size screens.

Sparse spectrum technique have been known for several years, but did not see much use in applications so far. SS produces unbiased non-periodic samples, and is capable to simulate phase with zero inner scale, and arbitrary large outer scale without additional corrections [9]. The number and location of spatial points are uncoupled from the number of spectral components. Since the number of spectral components does not grow with the screen size, SS computation times per screen grow slower than ones for the DFT-based techniques. Generation of SS random wave vectors with probability distributions matching the target spectrum can be challenging, but we found that about 500 SS components are sufficient to accurately approximate the Von Karman spectrum by uniform distributions at individual domains of spectral partition. After that, SS algorithm proved to be faster than both DFT and PWD for 1-megapixel and larger phase screens.

Recently proposed PWD technique [7] and its subharmonics-augmented version attempt to combine FFT computation speed with the randomized sampling essential to SS by using random wave vectors uniformly distributed at the rectangular segments of the spatial frequencies plane. Similar to the DFT, PWD can only generate phase on the rectangular spatial grid, and requires at least as many spectral components as there are spatial points. PWD-SH is capable of generating unbiased phase samples as long as the full support of the phase spectrum is captured by combination of the DFT and SH spectral domains. While even pure PWD provides an adequate coverage of the low-frequency spectral domain, the capture of the high-frequency part of the spectrum may require the impractically large number of DFT terms. Authors of [7] implemented an ad hoc unphysical addition of the white noise to the PWD-SH samples to alleviate this deficiency. Our calculation showed that PWD algorithm is about two times slower than the DFT, and that it is inferior to both SS and SU algorithms for screens larger than 0.25 megapixels. Additionally, the PWD performance deteriorates substantially for complex spectral models.

We introduced here a new version of the sparse spectrum approach to generation of homogeneous random fields – Sparse Uniform (SU) technique. As any sparse spectrum method, SU does not attempt to provide a dense coverage of the spectral plane for each sample, but uses a finite number of spectral components. Similar to the PWD, individual spectral components have uniform distributions of wave vectors over certain domains in the spectral plane and random complex amplitudes with compound distributions depending on the phase spectrum. It is straightforward to use SU algorithm for complicated spectral models, but this comes at the expense of additional calculations inside the Monte-Carlo loop. Still, computational times for SU and SS are very similar in our trails. SU generates unbiased phase samples, is capable of handling spectra with infinite outer scale, and is faster than the PWD-SH for screens larger than 0.25 MP.

Sample Matlab codes for all four algorithms are available in [13].

## 7. Appendix A: Non-biased property of SU technique.

Consider some domain $\Omega$ at the wave vector plane. According to Eq. (3) contribution of this domain to the structure function is

$$D_\Omega(\mathbf{r}) = 2 \iint_\Omega d^2\kappa \Phi(\mathbf{\kappa})[1 - \cos(\mathbf{\kappa} \cdot \mathbf{r})]. \tag{A1}$$

Consider single complex spectral component

$$\psi^{(\Omega)}(\mathbf{r}) = a^{(\Omega)} \exp(i\widetilde{\mathbf{\kappa}} \cdot \mathbf{r}) \tag{A2}$$

with uniformly distributed in $\Omega$ random wave vector $\widetilde{\mathbf{\kappa}}$ and compound complex amplitude

$$a^{(\Omega)} = (\alpha + i\beta)\sqrt{\Phi(\widetilde{\mathbf{\kappa}})S_\Omega} \tag{A3}$$

where $\alpha$ and $\beta$, are i.i.d. $N(0,1)$ random variables, and $S_\Omega$ is the area of $\Omega$. The complex phase structure function

$$\begin{aligned}\left\langle \left|\psi^{(\Omega)}(\mathbf{R}+\mathbf{r}) - \psi^{(\Omega)}(\mathbf{R})\right|^2\right\rangle_{\alpha,\beta,\widetilde{\mathbf{\kappa}}} &= 2\left\langle\left\langle a^{(\Omega)}a^{(\Omega)*}\right\rangle_{\alpha,\beta}[1 - \exp(i\widetilde{\mathbf{\kappa}} \cdot \mathbf{r})]\right\rangle_{\widetilde{\mathbf{\kappa}}} \\ &= 4S_\Omega \left\langle \Phi(\widetilde{\mathbf{\kappa}})[1 - \exp(i\widetilde{\mathbf{\kappa}} \cdot \mathbf{r})]\right\rangle_{\widetilde{\mathbf{\kappa}}} = 4\iint_\Omega d^2\kappa \Phi(\widetilde{\mathbf{\kappa}})[1 - \exp(i\widetilde{\mathbf{\kappa}} \cdot \mathbf{r})]\end{aligned} \tag{A4}$$

properly reproduces, after accounting for real phases, the sought contribution of the $\Omega$ domain to the structure function.

### References

1. J. Martin and S. M. Flatte, "Intensity images and statistics from numerical simulation of wave propagation in 3-D random media," Appl. Opt. **27**, 2111–2126 (1988).
2. R. Frehlich, "Simulation of laser propagation in a turbulent atmosphere," Appl. Opt. **39**, 393–397 (2000).

3. M. Charnotskii, "Sparse spectrum model for a turbulent phase," J. Opt. Soc. Am. A **30**, 479–488 (2013).
4. B. L. McGlamery, "Computer simulation studies of compensation of turbulence degraded images," Proc. SPIE **74**, 225–233 (1976).
5. R. Lane, A. Glindemann, and J. Dainty, "Simulation of a Kolmogorov phase screen," Waves Random Media **2**, 209–224 (1992).
6. M. Charnotskii, "Statistics of the sparse spectrum turbulent phase," J. Opt. Soc. Am. A **30**, 2455–2465 (2013).
7. D. A. Paulson, C. Wu, and C. C. Davis, "Randomized spectral sampling for efficient simulation of laser propagation through optical turbulence," J. Opt. Soc. Am. B **36**, 3249–3262 (2019)
8. V. I. Tatarskii, *Wave Propagation in a Turbulent Medium* (McGraw-Hill, 1961).
9. M. Charnotskii, "Wave propagation modeling with non-Markov phase screens," J. Opt. Soc. Am. A **33**, 561–569 (2016).
10. D. Voelz, E. Wijerathna, A. Muschinski, and Xifeng Xiao, "Computer simulations of optical turbulence in the weak- and strong scattering regime: angle-of-arrival fluctuations obtained from ray optics and wave optics," Opt. Eng. **57**, 104102 (2018).
11. L. Andrews, S. Vester, and C. Richardson, "Analytic expressions for the wave structure function based on a bump spectral model for refractive index fluctuations," J. Mod. Opt. **40**, 931–938 (1993).
12. V.V. Mikishov and V. I. Nikishov, "Spectrum of turbulent fluctuations of the sea-water refractive index," Int. J. Fluid. Mech. Res. **27**, 82 – 98 (2000).
13. M. Charnotskii, "Four code examples for turbulent phase screens simulation," figshare. Software. https://doi.org/10.6084/m9.figshare.10565714.v2.